\documentstyle[preprint,aps,floats]{revtex}
\begin{document}

\preprint{\tighten \vbox{\hbox{NSF-ITP-00-57}
  \hbox{hep-th/0006114}}}
\draft
\title{One-loop Shift in Noncommutative 
Chern-Simons Coupling}
\author{Guang-Hong Chen$^{1,2}$ and 
Yong-Shi Wu$^{1}$}
\address{1. Department of Physics, 
University of Utah,\\
Salt Lake City, Utah  84112\\
2. Institute for Theoretical Physics,\\
University of California, Santa Barbara, 
CA 93106\\
\vspace{.5cm} 
{\tt ghchen@physics.utah.edu \\
     wu@physics.utah.edu}
}

\maketitle
\begin{abstract}
\baselineskip=0.90cm
In this paper we study the one-loop shift 
in the coupling constant in a noncommutative 
pure $U(N)$ Chern-Simons gauge theory in 
three dimensions. The one-loop shift is 
shown to be a constant proportional to $N$, 
independent of noncommutativity parameters, 
and non-vanishing for $U(1)$ theory. 
Possible physical and mathematical 
implications of this result are discussed.

\end{abstract}
%\pacs{PACS numbers: blar-blar-blar}
\newpage
 %\narrowtext

\section{Introduction}

Field theory (especially gauge field theory) 
on a noncommutative space (or spacetime) has  
attracted much interest recently 
\cite{filk,kra,martin,bigatti,seiberg,hay,suss,gomis,chu,german,armoni}. 
That such theory arises from string/M(atrix) 
theory \cite{matrix,witten} suggests 
that space or spacetime noncommutativity 
should be a general feature of quantum 
gravity for generic points deep inside the 
moduli space of M-theory. Moreover, being 
a natural deformation of usual quantum field 
theory, noncommutative field theory is of 
interests in its own right. A charge in 
the lowest Landau level in a strong magnetic 
field can be viewed as living in a 
noncommutative space, because the 
guiding-center coordinates of the charge 
are known not to commute. In this paper, 
we study noncommutative Chern-Simons (NCCS) 
field theory in 3 dimensions, which is a 
deformation of ordinary Chern-Simons (CS) 
theory and may have applications in planar 
condensed matter systems, especially in 
the quantum Hall systems. For simplicity, 
in this paper we mainly consider {\it pure} 
NCCS theory, with gauge group $U(N)$ and 
with no matter fields coupled to it. 

Three-dimensional noncommutative spacetime 
has coordinates satisfying 
\begin{equation}
 [x^\mu, x^\nu]=i\theta^{\mu\nu} 
\hspace{1cm} \mu, \nu = 0,1,2, 
\label{noncom}
\end{equation}
where $\theta^{\mu\nu}$ are antisymmetric 
and real parameters of dimension length 
squared. The action for a pure $U(N)$ 
CS theory on this space reads 
\begin{equation}
\label{nccs}
I_{CS}=-\frac{i\kappa}{4\pi}\int d^3x
\varepsilon^{\mu\nu\lambda}\mbox{Tr}(
A_{\mu}*\partial_{\nu}A_{\lambda}
+\frac{2}{3}A_{\mu}*A_{\nu}*A_{\lambda}).
\end{equation}
Here the dynamical field is the gauge 
potential gauge potential $A^{\mu}
= A^a_\mu T^a$, $T^a$ the generators
of the gauge group $G=U(N)$, normalized 
to $ \mbox{Tr}(T^a T^b)= - \delta^{ab}/2$ 
with $T^{0}=i/\sqrt{2N}$ for the $U(1)$ 
sector. $\kappa$ is the CS coupling, 
$\varepsilon^{\mu\nu\lambda}$ 
the totally antisymmetric tensor with 
$\varepsilon^{012}=1$. In the action 
(\ref{nccs}) we are using a 
representation, in which the coordinates 
$x^\mu$ are the same as usual, but the 
product of any two functions of $x^\mu$ 
is deformed to the Moyal star-product:
\begin{equation} 
\label{starproduct}
f*g(x)=e^{(i/2)\theta^{\mu\nu}\partial^{x}_{\mu}
\partial^{y}_{\nu}} f(x)g(y)|_{y=x}.
\label{star}
\end{equation}
The commutator in Eq. (\ref{noncom}) is 
understood as the Moyal bracket with respect 
to the star product:
\begin{equation}
 [x^{\mu},x^{\nu}]\equiv 
x^{\mu}*x^{\nu}-x^{\nu}*x^{\mu}.
\end{equation}
For applications to a system in the lowest 
Landau level, one considers only the spatial 
noncommutativity:
$\theta^{01}=\theta^{02}=0$ and 
$[x^{1},x^{2}] =i\theta$. 

It is obvious that if $\theta^{\mu\nu}=0$, 
the action (\ref{nccs}) reduces to that of 
ordinary pure CS theory in 3 dimensions 
\cite{early-cs}, which is known to be a 
topological quantum field theory
\cite{witten88}, with the partition 
function and the correlation functions 
of Wilson loops being topological 
invariants, independent of spacetime 
metric. Diagrammatically the ordinary CS 
theory is renormalizable \cite{deser-rao}. 
Many topological features can be probed 
in perturbation theory \cite{csw1}. One 
interesting result is the one-loop 
quantum shift of the non-Abelian CS 
coupling \cite{csw1}. However, ordinary 
pure Abelian CS theory has no such shift, 
though additional matter coupling does 
at the two-loop level\cite{ssw,csw2,csw3}. 
In this paper we will show that 
there is a non-vanishing one-loop shift 
in noncommutative CS coupling even if 
the gauge group is $U(1)$. This shift 
turns out to be a constant proportional 
to the integer $N$, independent of 
the noncommutativity parameters 
$\theta^{\mu\nu}$, and identical to the
one-loop shift in ordinary $SU(N)$ CS
theory when $N\geq 2$. Possible physical 
and mathematical implications of our 
results will be discussed. 

\section{Regularized Feynman Rules}

The action (\ref{nccs}) is invariant 
under the following infinitesimal 
gauge transformations:
\begin{equation}
\delta A_\mu = D_{\mu} \lambda 
\equiv \partial \lambda 
+ [A_\mu, \lambda].   
\label{gaugetr}
\end{equation}
To do perturbation theory, we follow 
the standard procedure of path integral 
quantization to establish the Feynman 
rules. The full, regularized action after 
gauge fixing in Euclidean spacetime reads
\begin{equation}
\label{full}
I_{tot}=I_{CS}+I_{YM}+I_{gf}+I_{gh}.
\end{equation}
Here we have added the noncommutative 
Yang-Mills (YM) term
\begin{equation}
\label{u1ym}
I_{YM}=-\frac{1}{2e^2}\int d^3x 
\mbox{Tr}(F_{\mu\nu}*F^{\mu\nu}),
\end{equation}
with the field strength $F_{\mu\nu}$ 
defined by
\begin{equation}
\label{strength}
F_{\mu\nu}=\partial_{\mu}A_{\nu}
-\partial_{\nu}A_{\mu}+[A_{\mu},A_{\nu}].
\end{equation}
This gauge invariant term in the action 
provides a higher-derivative regularization 
for the CS theory, since the YM coupling 
$e^2$ is of dimension of mass, which is 
used as a cut-off that is sent to infinity 
at the end of calculations. The third term 
in Eq. (\ref{full}) is the gauge fixing 
term:
\begin{equation}
\label{gaugefix}
I_{gf}=-\frac{1}{\alpha e^2}\int d^3x\mbox{Tr}
(\partial^{\mu}A_{\mu})^2,
\end{equation}
a linear, covariant gauge condition 
convenient for perturbation theory. 
In the following we are going to take 
the Landau gauge $\alpha=0$, which was 
known to have computational advantages 
in the infrared in ordinary CS theory 
\cite{deser-rao,csw3}. The last term 
$I_{gh}$ is the ghost action 
corresponding to the above gauge fixing: 
\begin{equation}
\label{ghostaction}
I_{gh}=\int d^3x \mbox{Tr} (\partial^{\mu}
\overline{c}
D_{\mu}c), 
\end{equation} 
where $c$ and $\overline{c}$ 
are the ghost and anti-ghost respectively. 

In ordinary Yang-Mills theory, to write down 
the Feynman rules, in addition to full action 
(\ref{full}),  one more thing one needs to 
know is the representation of the gauge group, 
since it determines the normalization of the 
group factors. In the following, we will mainly 
concentrate on $U(1)$ theory, and we will come to 
$U(N)$ case naturally after the explicit calculation 
for $U(1)$ case. In noncommutative $U(1)$ 
gauge theory, though the group factor is trivial,
what is {\em nontrivial} is the noncommutativity 
of the kernel in Fourier transform. Suppose we 
have two kernels of Fourier transform 
${\cal F}_{k}=e^{ik\cdot x}$ and 
${\cal F}_{p}=e^{ip\cdot x}$, by using  
Eq. (\ref{starproduct}), one can easily check 
the following commutator
\begin{equation}     
\label{kercom}
[{\cal F}_{k},{\cal F}_{p}]
=-2i\sin(\frac{\theta^{ij}}{2}p_{i}k_{j})
{\cal F}_{k+p}=2i\sin(
\frac{\theta}{2}k\wedge p) {\cal F}_{k+p}.
\end{equation}
where $k\wedge p\equiv k_\mu 
\theta^{\mu\nu} p_\nu$, and we have 
used Eq. (\ref{noncom}). After making 
Fourier transform of the action, one can 
immediately find out that this commutator plays 
exactly the same role as that of Lie commutators 
of  the gauge group in ordinary non-Abelian 
gauge theory. Therefore, we can establish the 
Feynman rules by following the same procedure 
as that of ordinary Yang-Mills theory, with the 
group structure constants, $f^{abc}$, being 
replaced by a momentum-dependent factor
\cite{kra,martin},  namely,
\begin{equation}
\label{structurefunction}
f^{abc}\rightarrow f^{k,p,k+p}
=\sqrt{2}i\sin(\frac{\theta}{2}k\wedge p),
\end{equation}
where $\sqrt{2}$ is due to the normalization of $T^{0}$.
With the help of this correspondence, we 
establish the following Feynman rules
for $U(1)$ NCCS theory:

(i) The gluon propagator: 
\begin{equation}
\label{gluon}
\Delta_{\mu\nu}(p)=\frac{4\pi}{\kappa}
\frac{m}{p^2(p^2+m^2)}
(m\varepsilon_{\mu\nu\rho}p^{\rho}
+\delta_{\mu\nu}p^2-p_{\mu}p_{\nu}).
\end{equation}
where $m= e^2 \kappa /4\pi$. At the end
of computations, we remove the cut-off 
by taking $e^2\to \infty$ or $m\to \infty$.
 
(ii) The ghost propagator:
\begin{equation}
\label{ghostprop}
\frac{1}{p^2}.
\end{equation}

(iii) The ghost-ghost-gluon vertex:

\begin{equation}
\label{ghghgl}
-\sqrt{2}q_{\nu}\sin\biggl[{\frac{\theta}{2}q\wedge p}\biggr].
\end{equation}

(iv) The three gluon vertex:

\begin{equation}
\label{threegluon}
\frac{\kappa}{4\pi}\frac{\sqrt{2}}{m}
\sin\biggl[{\frac{\theta}{2}p\wedge q}\biggr]
[m\varepsilon_{\mu\nu\rho}
-(r-q)_{\mu}\delta_{\nu\rho}
-(q-p)_{\rho}\delta_{\mu\nu}
-(p-r)_{\nu}\delta_{\rho\mu}].
\end{equation}

(v)The four-gluon vertex:

\begin{eqnarray}
\label{fourgluon}
\frac{\kappa}{4\pi}\frac{1}{m}
[f^{p,q,t}f^{r,s,t} 
(\delta_{\mu\rho}\delta_{\nu\sigma}
- \delta_{\mu\sigma}\delta_{\nu\rho})
&+&f^{r,q,t} f^{p,s,t}
(\delta_{\mu\rho}\delta_{\nu\sigma}
-\delta_{\mu\nu}\delta_{\sigma\rho})\\ \nonumber
&+&f^{s,q,t}f^{r,p,t} 
(\delta_{\mu\nu}\delta_{\rho\sigma}
-\delta_{\mu\sigma} \delta_{\nu\rho})].
\end{eqnarray}
where $p,q,r,s,t$ are incoming momenta and $f^{x,y,z}$ 
is given by Eq. (\ref{structurefunction}).

\section{Ward-Slavnov-Taylor Identities}

In ordinary gauge theories, Ward-Slavnov-Taylor 
(WST) identities play a very important role in 
renormalized perturbation theory. For renormalizable 
gauge theories, they are essentially manifestation 
of gauge invariance for the regularized and 
renormalized action (with counter terms included). 
Conversely, checking WST identities is essentially 
checking renormalizability and gauge invariance of 
the renormalized gauge theory. The same is true for 
noncommutative gauge theories. In the following we 
are going to check part of the Ward identities to 
assure gauge invariance, and to use part of them 
to simplify the calculations. 

Renormalizability of the theory requires that
the full inverse $A$-propagator and the full 
$AAA$-vertex are of the following form as 
the external momenta tend to zero
\begin{equation}
\Delta^{-1}_{\mu\nu}(k)\to \frac{\kappa}{4\pi}
Z_A \epsilon_{\mu\nu\lambda} k_{\lambda}
+Z'_A (k^2\delta_{\mu\nu}-k_{\mu}k_{\nu}),
\quad (k\to 0),
\label{ZAs}
\end{equation}
\begin{equation}
\Gamma_{\mu\nu\lambda}(p,q,r)
\to Z_g 
\epsilon_{\mu\nu\lambda},
\qquad (p,q,r \to 0).
\label{Zg}
\end{equation}
These equations defines the relevant 
renormalization constants $Z_A$, $Z'_A$ 
and $Z_g$. In next section, we will confirm
the validity of these equations for one-loop 
two-point and three-point functions, to 
verify renormalizability of the theory at 
the one-loop level. Similarly one 
can defined $Z_{gh}$ and $\tilde{Z_g}$,
the renormalization constants\footnote{
Here we would like to remind that all the 
renormalization constants we defined here are 
consistent with the conventions used in the 
refs. \cite{deser-rao,csw1,csw2,csw3},
while being the inverse of the standard ones 
used in many textbooks on quantum field 
theory.} for the ghost wave function and the 
$\overline{c}Ac$-vertex respectively, through 
the full ghost propagator and the full 
$\overline{c}Ac$-vertex
\begin{equation}
\tilde{\Delta}(p) \to 
\frac{1} {Z_{gh} p^2}, 
\qquad (p\to 0),
\label{Zgh}
\end{equation}
\begin{equation}
i \Gamma_{\mu}(p,q,r)
\to i \tilde{Z}_g  
p_{\lambda},
\qquad (p,q,r \to 0).
\label{tildeZg}
\end{equation}
In the next section we will see that
at least at one loop, these renormalization 
constants are in fact finite, i.e. they are 
independent of the cut-off. 

Assuming renormalizability and introducing 
the renormalized fields and renormalization 
constants, one can write the renormalized 
action as
\begin{equation}
I_{ren}=I_{CS}'+I_{gf}'+I_{gh}'
\label{ren}
\end{equation}
with 
\begin{equation}
\label{nccsr}
I_{CS}'=-\frac{i\kappa_r}{4\pi}
\int d^3x\varepsilon^{\mu\nu\lambda} 
\mbox{Tr} [Z_A^{-1} A^{(r)}_{\mu}\partial_{\nu}
A^{(r)}_{\lambda}
+\frac{2}{3} Z_g^{-1} A^{(r)}_{\mu}
*A^{(r)}_{\nu}*A^{(r)}_{\lambda}],
\end{equation}
\begin{equation}
\label{gaugefixr}
I_{gf}'=-\frac{1}{\alpha_{r}}\int d^3x \mbox{Tr}
(\partial^{\mu}A^{(r)}_{\mu})^2,
\end{equation}
\begin{equation}
\label{ghostr}
I_{gh}'=\int d^3x \mbox{Tr} (Z_{gh}^{-1} 
\partial^{\mu}\overline{c}^{(r)}\
\partial_{\mu}c^{(r)}
+\tilde{Z}_g^{-1} \partial^{\mu}
\overline{c}^{(r)} 
[A^{(r)}_{\mu},c^{(r)}]).
\end{equation}
The terms (\ref{nccsr}), (\ref{gaugefixr}) 
and (\ref{ghostr}) in the renormalized 
action (\ref{ren}) should be equal to, 
respectively, the corresponding terms 
(\ref{nccs}), (\ref{gaugefix}) and 
(\ref{ghostaction}) in the original action 
(\ref{full}), if they are expressed in 
terms of the bare fields through
\footnote{The relation between $A_\mu$
and $A_\mu^{(r)}$, though looks unusual, 
is appropriate for the CS theory. See e.g. 
Refs. \cite{deser-rao,csw1,csw3}.} 
\begin{equation}
\label{fieldr}
A_{\mu}=\frac{Z_A}{Z_g} A^{(r)}_{\mu},
\quad c=Z_{gh}^{-1/2} c^{(r)}, 
\quad \overline{c}=Z_{gh}^{-1/2}
\overline{c}^{(r)}.
\end{equation}
This requires that the renormalized CS
coupling be related to the bare one by
\begin{equation}
\kappa_r = \frac{Z_A^3}{Z_g^2} \kappa \, ,
\label{kappar}
\end{equation}
and that the following WST identity be
true for the renormalization constants: 
\begin{equation}
\label{Ward}
\frac{Z_A}{Z_{gh}}=\frac{Z_g}{\tilde{Z}_g}.
\end{equation}
As easy to check, the WST identities 
guarantee that the renormalized actions
(\ref{nccsr}) and (\ref{ghostr}) are 
gauge invariant.

\section{One-loop renormaliztion in $U(1)$ 
theory}

In this section, we first study at the 
one-loop level the renormalization of 
$U(1)$ NCCS theory, in particular the shift 
in the Chern-Simons coupling $\kappa$. 

Let us start with the ghost self-energy. It
contains a planar diagram contribution 
%$\tilde{\Pi}^{(1)}_{p}$ and a non-planar diagram 
%contribution $\tilde{\Pi}^{(1)}_{np}$
\begin{equation}
 \label{ghostselfpl}
\tilde{\Pi}^{(1)}_{p}=\frac{4\pi}{\kappa}
\frac{m}{p^2} \int\frac{d^3k}{(2\pi)^3}
\frac{k^2p^2-(k\cdot p)^2}{k^2(k^2+m^2)(k+p)^2},
\end{equation}
and a non-planar diagram contribution 
%$\tilde{\Pi}^{(1)}_{np}$
\begin{equation}
 \label{ghostselfnpl}
\tilde{\Pi}^{(1)}_{np}
=\frac{4\pi}{\kappa}\frac{m}{p^2}
\int\frac{d^3k}{(2\pi)^3} 
\frac{k^2p^2-(k\cdot p)^2}
{k^2(k^2+m^2)(k+p)^2}e^{i\theta k\wedge p}.
\end{equation}

The integral (\ref{ghostselfpl}) is finite, with the
leading term proportional to $1/|m|$. Therefore, in 
the large $|m|$ limit, we get the contribution to 
the ghost self-energy as
\begin{equation}
\label{ghoren}
\tilde{\Pi}^{(1)}_{p}=-\frac{2}{3\kappa}
\mbox{sgn}(\kappa).
\end{equation}
On the other hand, the non-planar diagram 
contribution is evaluated to be
\begin{equation}
\label{ghonp}
\tilde{\Pi}^{(1)}_{np}=-\frac{2}{3\kappa}
\mbox{sgn}(\kappa) f(p\theta|m|),
\end{equation}
where the function $f(x)$ is defined by
\begin{equation}
\label{fun}
f(x)=\frac{1}{x}\int_0^{x} dy(3y^{1/2}
-y^{3/2})K_{1/2}(y)=\sqrt{2\pi}
\frac{(1+x)e^{-x}-1}{2x}.
\end{equation}
Since $f(x)\rightarrow 0$ as $x\rightarrow\infty$, 
the non-planar diagram does not contribute to the
ghost self-energy in the limit $|m|\rightarrow\infty$. 
Therefore, the ghost self-energy correction at the 
one loop level is finite and independent of external 
momentum $p$. Correspondingly, we get the 
one-loop ghost wave function renormalization 
constant
\begin{equation}
\label{ghwave}
Z^{(1)}_{gh}=1-\frac{2}{3\kappa}
\mbox{sgn}(\kappa).
\end{equation}

To calculate the gluon self energy 
$\Pi^{(1)}_{\mu\nu}(p)$, we decompose it into 
the following structure
\begin{equation}
\label{gluonself}
\Pi^{(1)}_{\mu\nu}(p)=\frac{1}{m}\Pi^{(1)}_{e}
(\delta_{\mu\nu}p^2-p_{\mu}p_{\nu})
+\frac{\kappa}{4\pi}\Pi^{(1)}_{o}(p)
\varepsilon_{\mu\nu\lambda}p^{\lambda}
\end{equation}
Since only the gluon loop diagram has an odd 
number of $\varepsilon$ tensors, 
$\Pi^{(1)}_{o}(p)$ receives a nonzero 
contribution only from the gluon loop diagram
Fig. 2a.  In contrast, $\Pi^{(1)}_{e}$ picks up 
contributions from the tadpole diagram (Fig. 2c), 
the ghost loop diagram Fig. 2b, as well as from 
the gluon diagram Fig. 2a with an even number 
of $\varepsilon$ tensors. Contracting 
$\Pi^{(1)}_{\mu\nu}$ with 
$\kappa/4\pi(\varepsilon_{\mu\nu\lambda} 
p^{\lambda}/2p^2)$ and $m\delta_{\mu\nu}/2p^2$, 
we obtain $\Pi^{(1)}_{o}(p)$ and $\Pi^{(1)}_{e}(p)$:
\begin{equation}
\label{pio}
\Pi^{(1)}_{o}=\frac{4\pi}{\kappa}\frac{2m}{p^2}
\int\frac{d^3k}{(2\pi)^3}[\sin^2(\theta/2k\wedge p)]
\frac{[k^2p^2-(k\cdot p)^2][5k^2+5(k\cdot p)
+4p^2+2m^2]}{k^2(k+p)^2(k^2+m^2)[(k+p)^2+m^2]},
\end{equation}
and
\begin{equation}
\label{pie}
\Pi^{(1)}_{e}=-\frac{m}{2p^2}\biggl[
\int\frac{d^3k}{(2\pi)^3}[\sin^2(\theta/2k\wedge p)]
\frac{N_{e}(p,k)}{k^2(k+p)^2(k^2+m^2)[(k+p)^2+m^2]}
+\frac{5m}{3\pi}\biggr].
\end{equation}
where 
\begin{eqnarray}
\label{nume}
N_{e}(p,k)=6k^6+18k^4(k\cdot p)+20k^4p^2
+22k^2(k\cdot p)^2p^2-12(k\cdot p)^3 
\\ \nonumber
+9k^2p^4-7(k\cdot p)^2 p^2+m^2[2k^4+4k^2(k\cdot p)
+k^2p^2+(k\cdot p)^2].
\end{eqnarray}

At this point we would like to comment that 
the structures shown in the above results are 
similar to those in ordinary non-Abelian 
Chern-Simons gauge theory in $2+1$ 
dimensions, although here we are dealing with
the $U(1)$ case. Still they are different in the 
following two aspects. The first is that we have 
non-planar diagram contributions due to the 
oscillating factor $4\sin^2(\theta/2 k\wedge p)$. 
The second is that the value of the tadpole 
contribution changes (see the second term in (\ref{pie})),
 the reason being that 
one of the terms in the four-gluon vertex vanishes 
due to the fact $\sin(\theta/2 p\wedge p)=0$ by 
using the Feynman rule (\ref{fourgluon}).  

The integral in Eq. (\ref{pio}) is finite. To calculate 
it, again we separate it into planar and 
non-planar contributions. The calculation of the planar 
contribution is standard. Taking $|m|\rightarrow \infty$, 
we obtain
\begin{equation}
\label{piopl}
\Pi^{(1)}_{o,p}=\frac{7}{3\kappa}\mbox{sgn}(\kappa).
\end{equation}

By using Feynman parameterization, we can rewrite 
the corresponding non-planar contribution as follows:
\begin{equation}
\label{pionp}
\Pi^{(1)}_{o,np}=-2p^2\int_{0}^{1}dx\int_{0}^{x}dy
\int_{0}^{y}dz\{5I_2(\nu)+[5p^2(1+y-x-z)(y-x-z)
+4p^2+2m^2]I_{1}(\nu)\},
\end{equation}
where the argument $\nu$ in the functions $I_{1}$ 
and $I_{2}$ is defined as
\begin{equation}
\label{argu}
\nu=\sqrt{p^2(1+y-x-z)+m^2(1-y)}.
\end{equation}

The functions $I_1$ and $I_2$ are defined as
\begin{equation}
\label{I1}
I_1=\frac{\pi^{3/2}}{2}\biggl(\frac{\theta p}{2}
\biggr)^{3/2}[\nu^{-3/2}K_{3/2}(\nu\theta p)
-\frac{\theta p}{3}\nu^{-1/2}K_{1/2}(\nu\theta p)],
\end{equation}
 and
\begin{equation}
I_2=\frac{\pi^{3/2}}{12}\biggl(
\frac{\theta p}{2}\biggr)^{1/2}[15\nu^{-1/2}
K_{1/2}(\nu\theta p)-20\biggl(\frac{\theta p}{2}
\biggr)^{3/2}\nu^{1/2}K_{1/2}(\nu\theta p)
+4\biggl(\frac{\theta p}{2}\biggr)^{5/2}
\nu^{3/2}K_{3/2}(\nu\theta p)],
\end{equation}
where $K_{\beta}(x)$ is the modified Bessel 
function. It has an exponentially decay profile. 
In the limit  $|m|\rightarrow\infty$, we see that 
the integrand in Eq. (\ref{pionp}) vanishes, 
therefore the non-planar diagram does not 
contribute to $\Pi^{(1)}_{o}$. Namely,
\begin{equation}
\label{pionp2}
\Pi^{(1)}_{o,np}=0.
\end{equation}
 Thus, we 
get 
\begin{equation}
\label{piofull}
\Pi^{(1)}_{o}=\frac{7}{3\kappa}
\mbox{sgn}(\kappa).
\end{equation}

Similar analysis can be applied to the integral 
(\ref{pie}). It turns out that the integral is 
finite as $|m|\rightarrow\infty$. Therefore, the 
photon wave function renormalization constant is 
\begin{equation}
\label{za}
Z^{(1)}_A=1+\Pi^{(1)}_{o}=1+\frac{7}{3\kappa}
\mbox{sgn}(\kappa).
\end{equation}

Furthermore, we study the one loop corrections to 
the vertex $\overline{c}Ac$, we show that the one 
loop correction vanishes as $|m|\rightarrow\infty$. 
The reason is that the one-$\varepsilon$ term of 
Fig. 2(i) cancels against the three-$\varepsilon$ 
term of Fig. 2(h); the one-$\varepsilon$ term of 
Fig. 2(h) goes to zero; the non-$\varepsilon$ terms 
in the two diagrams cancel each other as well. 
Finally, we get the renormalization for the 
$\overline{c}Ac$ vertex 
\begin{equation}
\label{zcac}
\tilde{Z}^{(1)}_g=1.
\end{equation}

After we extract $Z^{(1)}_{gh}$, $Z^{(1)}_{A}$, 
and $\tilde{Z}^{(1)}_{g}$, we can employ the 
WST identity (\ref{Ward}) established in the
previous section to get the three-gluon vertex 
renormalization constant: 
\begin{equation}
\label{zg}
Z^{(1)}_{g}=\frac{Z^{(1)}_{A}}{Z^{(1)}_{gh}}
\tilde{Z}^{(1)}_{g}=
1+\frac{3}{\kappa}\mbox{sgn}(\kappa), 
\end{equation}
where we have worked up to the first order in
$1/\kappa$, consistent with one-loop 
perturbative theory. 

Now we have shown that all renormalization 
constants at the one loop level are finite, 
so the one-loop beta function vanishes, as
in ordinary CS theory. Substituting the 
renormalization constants $Z^{(1)}_{A}$ and 
$Z^{(1)}_{g}$ in the definition of the 
renormalized CS coupling $\kappa^{(1)}_r$, 
we have
\begin{equation}
\label{massshift}
\kappa^{(1)}_r=\kappa+\; 
\mbox{sgn}(\kappa).
\end{equation}
This is the main result of the present 
paper. The second term is the desired 
one-loop shift in the $U(1)$ Chern-Simons 
coupling. Note that the shift is just
the unity in our normalization for the 
coupling. Note that it is independent 
both of the noncommutativity parameters 
$\theta^{\mu\nu}$ and of the value of 
the bare coupling $\kappa$ except for 
its sign. Also recall that for ordinary
$U(1)$ CS theory, the one-loop shift 
vanishes in the same $F^2$ regularization.

\section{Generalization to $U(N)$}

In this section, we generalize the 
expression (\ref{massshift}) we obtained 
in last section for the one-loop shift 
of the CS coupling from $U(1)$ to $U(N)$.

It is known that for a noncommutative
gauge theory, the gauge group is 
restricted to be only $U(N)$; even 
$SU(N)$ is not allowed, because the 
closure of the Moyal commutator is 
violated \onlinecite{Matsu}. Another 
way to see this is that the three-gluon 
coupling in the action (\ref{nccs}) 
mixes the $U(1)$ gluon with the 
$SU(N)$ gluons. Thus, unlike 
ordinary gauge theory which allows 
the $U(1)$ and $SU(N)$ sectors to
have independent coupling constants,
in noncommutative theory $U(N)$ 
gauge invariance enforces the 
$U(1)$ and the $SU(N)$ gluons to 
share the same coupling constant.

For $U(N)$ noncommutative Yang-Mills 
(NCYM) theory, it is this distinct 
feature that makes the coupling in 
the $U(1)$ sector runs in the same 
way as that in the $SU(N)$ sector, 
as verified in a recent explicit 
calculation \onlinecite{armoni}. Before 
this calculation was done, a beautiful 
proof without doing any new calculation 
had been given in ref. \onlinecite{seiberg}  
for the statement that the one-loop 
beta-function of $U(N)$ NCYM can be 
simply read off from the known value 
of the ordinary $SU(N)$ Yang-Mills 
theory. This is because in NCYM the 
nonplanar one-loop $U(N)$ diagram 
contributes only to the $U(1)$ part 
of the theory. In the following we 
will apply the same trick to NCCS,
and derive the one-loop shift in
the CS coupling in the $U(N)$ theory 
without doing any new calculations.

Following ref. \onlinecite{seiberg}, 
let us consider the quadratic one-loop 
1PI effective action of the ordinary 
$U(N)$ CS theory, in which the $U(1)$
and $SU(N)$ sectors share the same
coupling constant, in the $F^2$ 
regularization. From the results 
in refs. \onlinecite{csw1,csw3} one 
infers that after removing the cut-off,
\begin{equation}
\Gamma_2^{(1)}(\theta=0)=
-\frac{i}{4\pi}\int d^3x
\varepsilon^{\mu\nu\lambda}
[ (\kappa+ N \mbox{sgn}(\kappa)) 
\mbox{Tr} (A_{\mu}
\partial_{\nu}A_{\lambda})
- \mbox{sgn}(\kappa) (\mbox{Tr} 
A_\mu)\partial_\nu (\mbox{Tr} 
A_\lambda) ].
\label{1PIo}
\end{equation}
The coefficient $\kappa+ N\mbox{sgn}(\kappa)$ 
in the first term is read off 
from the known one-loop shift 
in ordinary $SU(N)$ CS theory
\onlinecite{csw1,csw3}, while the 
existence of the second term 
is due to the necessity for 
cancelling the $U(1)$ part in 
the first term, since we know
there is no one-loop shift in 
the $U(1)$ coupling constant.
Also it is easy to check that
the second term is the only
nonplanar contribution to
$\Gamma_2^{(1)}$, coming 
from the nonplanar part of 
diagrams like Fig. 2 (a)-(c). 
 
Now let us turn on nonzero 
$\theta_{\mu\nu}$. The planar
contributions to $\Gamma_2^{(1)}$
are known to be the same as in 
ordinary theory \onlinecite{bigatti,seiberg}, 
while the nonplanar diagrams are 
suppressed to zero by an extra rapidly 
oscillating phase factor. Our
result (\ref{pionp2}) in last 
section verifies the latter by   
explicit computation. Thus, with 
the second term put to zero, one
reads from eq. (\ref{1PIo}) that
in $U(N)$ NCCS,
\begin{equation}
\Gamma_2^{(1)}(\theta\neq 0)=
-\frac{i}{4\pi}\int d^3x
\varepsilon^{\mu\nu\lambda}
[\kappa+ N \mbox{sgn}(\kappa)] 
\mbox{Tr} (A_{\mu}\partial_{\nu}
A_{\lambda}).
\label{1PInc}
\end{equation}
Therefore, without any new calculation, 
we infer that the one-loop shift in the
coupling for $U(N)$ NCCS is
\begin{eqnarray}
\label{unshift}
\kappa_r
%&=&\kappa+\frac{1}{2}
%C_2(G)\mbox{sgn}(\kappa) \nonumber\\
&=&\kappa+N\mbox{sgn}(\kappa).
\end{eqnarray} 
For the $U(1)$ case, we plug $N=1$ 
in Eq. (\ref{unshift}), reproducing
exactly Eq. (\ref{massshift}) that
we have obtained by explicit 
calculation in last section. 

In summary, the one-loop shift of the
CS coupling in $U(N)$ NCCS is the same 
as that in ordinary $SU(N)$ CS theory
for $N\geq 2$, while for the $U(1)$
case it gives a non-vanishing value
in contrast to ordinary CS theory.

\section{Conclusions and Discussions}

The induced CS coupling by fermionic 
fields in noncommutative quantum 
electrodynamics in 3 dimensions has 
been studied in ref. \onlinecite{chu}.  
In this paper we have considered instead
 a pure CS theory without matter, 
and have studied the one-loop quantum 
correction to the coupling constant 
due to self-interactions of the gauge 
bosons that arise from spacetime 
coordinate noncommutativity. 

First of all, the renormalization 
constants we have calculated at the 
one loop level are finite, showing 
that the beta function at this level 
vanishes. Moreover all 
renormalization constants, including 
the one-loop shift in the CS coupling 
are shown to be independent of 
the noncommutativity parameters. 
This is a bit surprising, since the 
spacetime noncommutativity parameters 
appear in the Lagrangian of pure CS 
theory explicitly. This adds explicit 
evidence to a theorem proved in Ref. 
\onlinecite{german} that one-loop results 
in noncommutative CS theory are all 
independent of spacetime noncommutativity.
It would be interesting to see whether 
the same independence remains true at 
higher orders. Our conjectured 
answer is "yes". In other words, we 
conjecture that {\it noncommutative CS 
theory is a ``deformed" topological 
field theory}, in the sense that the 
partition function and correlation 
functions are topological invariants, 
independent of both metric and 
noncommutativity parameters.  

Furthermore, we have shown that the 
one-loop shift in the coupling 
constant does not vanish in a 
pure $U(1)$ NCCS theory. This arises 
as a consequence of spacetime coordinate
noncommutativity, since in ordinary 
$U(1)$ CS theory there is no quantum 
shift in the coupling constant at all. 
We notice two features of our result 
(\ref{massshift}): 1) it is independent 
of the bare coupling $\kappa$, except
for its sign; 2) it is the simplest 
integer, the unity, independent of the 
noncommutatvity parameters. Therefore 
this result is not smooth in the limit 
$\theta_{\mu\nu}\to 0$. 

Finally, we have shown that the one-loop 
shift of the $U(N)$ NCCS coupling is
the integer $N$ that characterizes the 
gauge group $U(N)$, exactly the same as 
that in ordinary $SU(N)$ CS theory for 
$N\geq 2$. In ordinary CS theory the 
quantization of the shift was interpreted 
\onlinecite{csw1,csw3} as being consistent 
with the topological quantization of 
the non-abelian CS coupling. The latter 
is known \onlinecite{early-cs} to result 
from the topological fact that a large 
gauge transformation changes the CS 
action by a value proportional to the 
integer winding number of the gauge 
transformation, viewed as a map from
the (compactified) spacetime to the
gauge group. Whether the topological 
quantization of the CS coupling remains 
true in the noncommutative theory is 
not clear at all at this moment. 
However, our result (\ref{unshift}) 
shows that the one-loop shift is still 
quantized in the noncommutative case. 
If one reverses the logic in the above 
reasoning for the ordinary CS theory, 
this result seems to indicate that 
possibly in noncommutative geometry 
there should be a counterpart of the 
concept of the usual winding number 
that remains integer-valued, and 
that the NCCS coupling should satisfy 
a similar topological quantization.

Here we would like to mention that 
the one-loop shift in the $U(1)$ NCCS 
coupling depends on the regularization 
used. Our result (\ref{massshift}) was 
obtained in the $F^2$ regularization, 
in which the Yang-Mills term was 
added to the action and its coefficient 
was taken as cut-off. If we had used 
dimensional regularization, the shift 
would be zero. This situation is not 
surprising, completely similar to the 
well-known situations for ordinary 
non-Abelian CS theory; see e.g. 
Refs. \onlinecite{csw1,csw3}. In our 
opinion, the $F^2$ regularization is
more physical, in the sense that
the $F^2$ term may naturally appear
in realistic planar systems.

In field theory on ordinary spacetime, the 
$U(1)$ CS coupling (CS coefficient) is 
known to have several interesting physical 
meanings, when the CS gauge field couples 
to various fields. For example, when there 
is a Maxwell term in the action, $\kappa$ 
is related to the topological mass of the 
CS photon \onlinecite{early-cs}. When there 
are matter fields coupled to the CS field, 
the CS coefficient $\kappa$ will give rise to 
fractional (exchange) statistics for matter 
field quanta \onlinecite{fes}. Finally a 
well-known folklore in the community is that an 
effective CS coupling for the electromagnetic 
field indicates the Hall effect, with the 
effective CS coefficient directly related 
to the Hall conductance. We expect all these 
physical interpretations should be 
generalizable to nocommutative spacetime. 
A systematic inverstigation of NCCS coupled 
to matter fields (both fermionic and bosonic) 
will be published elsewhere \onlinecite{chen-wu}.

\acknowledgments

One of us, GHC, thanks the Institute for 
Theoretical Physics, University of California 
at Santa Barbara, for an ITP Graduate 
Fellowship, and for the warm hospitality he 
receives during his stay. This research was 
supported in part by the National Science 
Foundation under Grants No. PHY-9407194(GHC) and 
PHY-9907701(YSW).

{\bf Notice added:} After the paper was put on the 
e-print archive, we are informed by e-mail
from Soo-jong Rey that he has obtained the 
same result (in agreement with ours) on $U(1)$ 
NCCS theory.

%\end{flushleft}

\newpage
\begin{figure}
\caption{Feynman rules. }
\end{figure}
\begin{figure}
\caption{ One-loop Feynman diagrams in pure Chern-Simons theory 
(solid line-gluon;dashed line-ghost: (a)-(c) gluon self-energy, 
(d) ghost self-energy, (e)-(g) three-gluon vertex, (h), (i) 
ghost-gluon vertex. }
\end {figure}

\end{document}